\documentclass[aps,pra,reprint,amsmath,amssymb,floatfix,longbibliography,superscriptaddress]{revtex4-2}

\usepackage{graphicx}
\usepackage{bm}

\newcommand{\E}{\mathrm{E}}

\begin{document}
	
	\title{Constraints on the Origin of Universal \boldmath$1/f^2$ Photon-Count Spectra at Baseband}
	
	\author{Malvin Carl Teich}
	\email[Contact author: ]{teich@bu.edu}
	\affiliation{Department of Electrical \& Computer Engineering, Department of Physics, and Photonics Center, Boston University, Boston, Massachusetts 02215, USA}
	\affiliation{Department of Electrical Engineering, Columbia University, New York, New York 10027, USA}
	
	\author{Steven B. Lowen}
	\altaffiliation[Present address: ]{Insulet Corporation, Acton, MA 01720, USA}
	\affiliation{Department of Electrical \& Computer Engineering and Photonics Center, Boston University, Boston, Massachusetts 02215, USA}

	\date{\today}
	
\begin{abstract}
	A series of 11.6-d-duration photon-counting experiments, employing a broad variety of light sources with different statistical properties and optical spectra, were carried out over a 1.5-yr period.
	All of the photon-count spectra at baseband followed a common $1/f^2$ form over the frequency range $1 \times 10^{-6} \leqslant f \leqslant 5 \times 10^{-4}$~Hz, corresponding to a timescale range $33$~min $ \leqslant T_f \leqslant 11.6$~d, where $T_f \equiv 1/f$. The lower and upper timescale limits were established by the photodetector noise floor and the duration of the individual experiments, respectively.
	Unlike ordinary Brownian motion, all of the measured photon-count sample paths exhibited irregular long-timescale fluctuations, with durations ranging from hours to days. It has been established that the photon-count spectra cannot plausibly be ascribed to fluctuations of the current, voltage, or temperature of the source or the detector, nor to technical sources of noise associated with the optical system or local environment. Although the physical origin of the photon-count fluctuations remains unresolved, several heterodox hypotheses are set forth. From a statistical point-of-view, the photon counts appear to follow a doubly stochastic Poisson process with a slowly varying random intensity. 
	Analogous experiments that rely on ionizing radiation and direct-conversion solid-state radiation detectors are proposed.
\end{abstract}

	\maketitle
	
\section{Introduction} \label{sec:intro}
 
In the course of conducting a series of long-duration optical-imaging experiments in a dedicated photonics laboratory environment that made use of a panoply of different optical sources, it became apparent that at extremely low frequencies the strength of the photon-count fluctuations increased markedly with decreasing frequency~\cite{mohan20PRR}. 

We proceeded to study these fluctuations in numerous experiments conducted continuously over a period of 1.5~yr. It became clear that all of the light sources we investigated, including light-emitting diodes, superluminescent diodes, laser diodes, incandescent sources, and betaluminescent emitters, exhibited photon-count fluctuations that manifested an inverse-square spectral form at frequencies that extended to $<1~\mu$Hz~\cite{mohan20PRR}.

The robust character of this inverse-square baseband spectral behavior stands in sharp contrast to the optical spectra of these sources, which, while differing markedly from each other, were all in the frequency range 
$\sim 10^{14}$~Hz. 
This paper establishes constraints on the physical mechanisms underlying these baseband fluctuations, whose origin remains unclear.

\subsection{Photon Counting} \label{ssec:PhotocountIntro}

Photon counting is a fundamental measurement technique used in experimental quantum optics.
Our understanding of light as a collection of discrete packets of energy harks back to the work of Planck in 1901~\cite{planck01} and Einstein in 1905~\cite{einstein05}. (The term \textit{photon}, introduced by Lewis in 1926~ \cite{lewis26}, came to be used to describe what Einstein had originally termed \textit{Lichtquant}.) Soon thereafter, in 1909, Campbell examined the statistical properties of photon arrivals, specifically the random number of photons registered in sequential fixed counting windows of duration $T$~\cite{Campbell09b,campbell09a}. Campbell identified the result as a Poisson counting process, and specified its mean and variance. 

In 1918, Schottky examined the homogeneous Poisson point process (HPP) itself and chose the appellation \textit{shot effect} to describe its linearly filtered version~\cite{schottky18}. 
A fundamental property of the HPP is that it is memoryless, so that 
knowledge of the entire history and future of a given realization thereof yields no information about the behavior of the process at the present.
The HPP is characterized by a single constant
quantity, its rate $\mu$, which is the number of events expected to occur in a unit counting window.
Other point processes do not share this memoryless
property and therefore cannot be described in terms of a
constant rate.

The statistical independence of HPP photon registrations ensures that the photon counting process has a Poisson distribution~\cite{Poisson1837,	vonBortkiewicz1898,QuineSeneta1987}.
The probability of detecting $n$ photons in a set of consecutive counting windows of duration $T$, across a data set collected over a time $L$, may be written as~\cite[Sec.~13.2C]{saleh19}
\begin{equation}
	p(n; \mu T) = \frac{(\mu T)^n\, e^{-\mu T}}{n!}\,.
	%p(n; \mu T) = (\mu T)^n\, e^{-\mu T}/n!\,.
\end{equation}

\subsection{Photon-Count Fluctuations} \label{ssec:PhotcountFluct}

The photon-count fluctuations associated with light~\cite{mandel1959} are important in many physical and biological processes. Indeed, these fluctuations impact the design of experiments in fields that range from optical coherence tomography~\cite{mohan08,mohan09,mohan10} to gravitational-wave detection~\cite{abbott16,abac2026,payne2026}.

A substantial literature exists that details the  photon-counting statistics and coherence properties of various sources of light~\cite{saleh78,perina85,teich88,lowen05}. Although some forms of light, such as {\v C}erenkov radiation from a random stream of charged particles, exhibit fractal power-law spectra of the form $1/f^\alpha$ ($0 < \alpha < 2$) at baseband~\cite{lowen91pra,lowen05}, the observation of nonfractal $1/f^2$  photon-count fluctuations at low frequencies is both unexpected and perplexing.

The five classes of light sources investigated in our long-duration photon-counting experiments~\cite{mohan20PRR} included three semiconductor-based light sources: light-emitting diodes (LEDs), superluminescent diodes (SLEDs), and laser diodes (LDs). These types of devices emit different forms of electron--hole recombination radiation, namely spontaneous emission, amplified spontaneous emission (ASE), and stimulated emission, respectively \cite[Secs.~18.1, 18.2E, and 18.4]{saleh19}. We also examined the light emitted from incandescent sources, which rely on a heated filament that emits blackbody radiation \cite[Sec.~4.8]{teich2025}. Finally, we studied a source that operates on the basis of phosphor luminescence initiated by $\beta$-particles emitted from tritium atoms ($^3$H) \cite[][Sec.~14.5A]{saleh19}. Two types of photodetection systems were employed: a single-photon avalanche diode (SPAD) module and a photomultiplier tube (PMT) and its associated circuitry~\cite[Secs.~19.4C and 19.1A, respectively]{saleh19}.

\subsection{Section Outline} \label{ssec:secout}

Section~\ref{Exp} reviews the study we reported in 2020~\cite{mohan20PRR}, which was focused on observing the second-order statistics of the long-timescale photon-count fluctuations associated with these five classes of light sources. As such, attention was directed toward the photon-count periodograms and scalograms, both of which exhibited square-law behavior. Photon-count $z$-score time traces were also examined --- they displayed unusual and unexpected fluctuations over time periods ranging from hours to days for all of the light sources studied, without exception~\cite[Fig.~6]{mohan20PRR}.
 
Sections~\ref{sec:ledsledldincand} and \ref{sec:betalum} examine low-frequency fluctuations observed in electroluminescence and incandescence, and in betaluminescence, respectively, while Secs.~\ref{Results:STPG}, \ref{sec:detectnoise}, and \ref{sec:technoise} analyze low-frequency source-temperature fluctuations, detector noise, and technical noise, respectively. Section~\ref{sec:technoiseexclude} summarizes putative sources of low-frequency fluctuations
to which the presence of the $1/f^2$ photon-count spectrum cannot plausibly be ascribed. 
Section~\ref{sec:statfluct} assesses the statistical characteristics of the photon-count fluctuations, while Secs.~\ref{sec:possorig} and \ref{sec:spec} consider possible  individual- and common-origin heterodox hypotheses, respectively. 
Experiments involving ionizing-radiation that could elucidate the origin of the low-frequency fluctuations are proposed in Sec.~\ref{suggest} and 
the conclusion is presented in Sec.~\ref{concl}.

\section{Observation of Source-Independent Low-Frequency Photon-Count Fluctuations} \label{Exp}

\begin{figure}[htbp!]
	\centering
	\includegraphics[width=\linewidth]{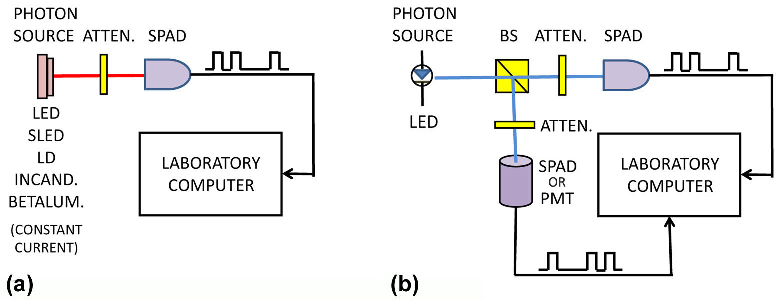}
	\caption{Experimental arrangements. (a) Five classes of photon sources were used in the experiments: LEDs (red, near-infrared, and blue), SLEDs (red and near-infrared), LDs (red and near-infrared), incandescent sources (white), and a betaluminescent source (green). Most, but not all, of the data discussed in this paper were collected from sources operated at constant current. The light from the source was attenuated and directed to a SPAD module. The source was located on an optical bench, at an arbitrary distance from the module, and the attenuator was adjusted such that the detected photon flux was in the vicinity of 200\,000~s$^{-1}$, which is more than an order of magnitude below its saturation value. The SPAD module served to detect individual photons and to convert them into idealized electronic pulses (as schematized in the figure), which were then sent to a laboratory computer for processing.
	(b) In the auxiliary experiments reported in Sec.~\ref{ssec:cc}, light from a red LED was split into two beams at a cube beamsplitter (BS), attenuated to an approximate photon flux of either 400\,000~s$^{-1}$ or 100\,000~s$^{-1}$ in each beam, and directed to a pair of SPAD modules.
	In the auxiliary experiments reported in Secs.~\ref{sec:detectnoise} and \ref{ssec:statfluctjoint}, light from a constant-current-driven blue LED was split into two beams at a cube beamsplitter (BS) and attenuated such that the photon flux in each beam was reduced to roughly 200\,000~s$^{-1}$. One of the beams was directed to a SPAD module while the other impinged on a PMT.
	(Adapted from \cite[Fig.~1]{mohan20PRR}.)} \label{setup}
\end{figure}

The experimental arrangements used in the baseband photon-count-fluctuation studies reported in~\cite{mohan20PRR} are illustrated in Fig.~\ref{setup}. 
In the simplest configuration [Fig.~\ref{setup}(a)], the light from the source, after passing through an attenuator,  was incident on a SPAD module that detected individual photons and converted them into idealized electronic pulses.
The numbers of pulses arriving at the SPAD module in successive counting windows of duration $T=5$~s were recorded.

\begin{figure*}[tb!]
	\centering
	\includegraphics[width=14cm]{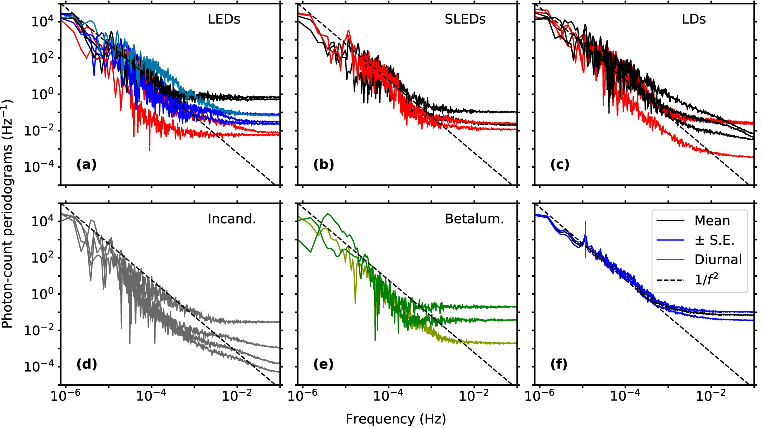}
	\caption{Photon-count periodograms plotted on doubly logarithmic coordinates. Individual normalized periodograms, $S_P(f)$ vs. frequency $f$, are plotted one atop the other for five classes of optical sources in panels (a)--(e). The colors in which the curves are plotted correspond to the colors of the light emitted by the individual devices (black and gray represent near-infrared and white incandescent light, respectively). The data displayed were collected using the SPAD module, except where indicated.
	The baseline noise for the SPAD becomes appreciable at $f \approx 5 \times 10^{-4}$~Hz (cutoff frequencies instigated by detector noise are far more limiting than those arising from detector response time and Nyquist sampling).
	(a)~LEDs (eight data sets) --- the cyan curve represents data collected with the PMT photodetection system; (b)~SLEDs (four data sets); (c)~LDs (six data sets); (d)~incandescent sources (four data sets); and (e)~a betaluminescent source (three data sets) --- the yellow-green curve represents data collected with the PMT photodetection system. Twenty-three of the 25  periodograms displayed in panels~(a)--(e) were derived from data sets of the same length (11.6~d); the other two curves, portrayed as green in panel~(e), had durations of 8.7~d and 9.3~d. (f)~Despite the fact that they were measured at different times and under different laboratory conditions, and that they corresponded to five different classes of sources, the 23 periodograms follow each other sufficiently closely that they can be averaged to improve statistical accuracy. The black curve represents the averaged periodogram $\overline{S}_P(f)$ vs. frequency $f$ and the two blue curves hugging it represent standard-error values ($\pm$~S.E.). The mean periodogram decreases approximately as $1/f^2$ (dashed line) over the frequency range $1.0 \times 10^{-6} \leqslant f \leqslant 5.0 \times 10^{-4}$~Hz. 
	The vertical red line at $1.16 \times 10^{-5}$~Hz corresponds to diurnal variations (1~d = 86\,400~s). (Adapted from \cite[Fig.~2]{mohan20PRR}.)
	\label{bothphotonpg} }
\end{figure*}

The counts were averaged over $\approx 83.33$~min before plotting, thereby rendering fluctuations with durations  shorter than $\approx~1$~h unresolvable. Fluctuations with durations longer than approximately 11.6 days cannot be observed since the maximum duration of an experiment was 11.6~d.

The optical sources, which made use of a range of operating principles, had a variety of optical spectra, apertures, radiation patterns, and coherence properties~\cite{mohan20PRR}.
The photon-counting measurements were conducted concomitantly with a series of temperature measurements, the details of which were reported separately~\cite{lowen20}.

Dual-detector configurations, such as that portrayed in 
Fig.~\ref{setup}(b), were employed in the experiments reported in Secs.~\ref{ssec:cc}, \ref{sec:detectnoise}, and \ref{ssec:statfluctjoint}.

Details of the laboratory configurations and photodetection systems used in the experiments are provided in~\cite{mohan20PRR}, as are descriptions of the source-temperature monitoring and control systems, data-analysis protocols, and computational procedures. 

The experiments were carried out in a dedicated laboratory at the Boston University Photonics Center, where we could control, restrict, and suspend climate control (heating and air conditioning), janitorial services, and access to personnel. Individual experimental runs, which typically lasted approximately 14 days, were conducted continuously over a period of 1.5 years, from May 2010 until December 2011.

\subsection{Photon-Count Periodograms} \label{Results:PCPG}

Photon-count periodograms for all five classes of optical sources are displayed in Fig.~\ref{bothphotonpg}.
The data were collected using the experimental configuration portrayed in Fig.~\ref{setup}(a) and the periodograms were computed using the procedure described in Sec.~II F of~\cite{mohan20PRR}. Figures~\ref{bothphotonpg}(a)--\ref{bothphotonpg}(e) represent collections of  individual periodograms for LEDs, SLEDs, LDs, incandescent sources, and a betaluminescent source, respectively.

\begin{figure*}[tb!]
	\centering
	\includegraphics[width=14cm]{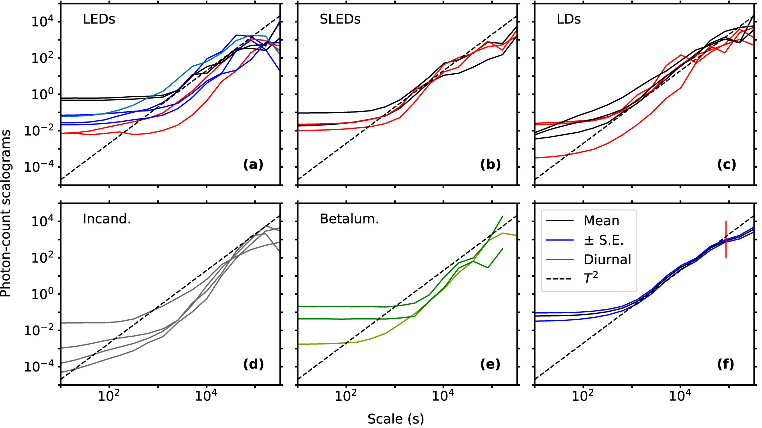}
	\caption{Photon-count scalograms plotted on doubly-logarithmic coordinates. Individual normalized Daubechies 4-tap wavelet variance curves, $A_P(T)$ vs. scale $T$~\cite{vetterli95,daubechies92,lowen05}, are plotted one atop the other for the five classes of optical sources in panels (a)--(e). The colors in which the curves are plotted correspond to the colors of the light emitted by the individual devices (black and gray represent near-infrared and white light, respectively). The data  presented were collected using the SPAD module, except where indicated. (a)~LEDs (8~data sets) --- the cyan curve represents data collected with the PMT photodetection system; (b)~SLEDs (4~data sets); (c)~LDs (6~data sets); (d)~incandescent sources (4~data sets); and (e)~a betaluminescent source  (3~data sets) --- the yellow-green curve represents data collected with the PMT photodetection system. Twenty-three of the 25 device scalograms displayed in panels~(a)--(e) were derived from data sets of the same length (11.6~d); the other two curves, portrayed as green in panel~(e), had durations of 8.7~d and 9.3~d. (f)~Despite the fact that they were measured at different times and under different laboratory conditions, and that they corresponded to five different classes of sources, the 23 scalograms follow each other sufficiently closely that they can be averaged to improve the statistical accuracy. The black curve represents the averaged Daubechies 4-tap wavelet variance $\overline{A}_P(T)$ vs. scale $T$ and the two blue curves hugging it represent standard-error values ($\pm$~S.E.). Diurnal variations are indicated by the vertical red line (1~d = $8.64 \times 10^4$~s).  (Adapted from~\cite[Fig.~5]{mohan20PRR}.)\label{bothphotonwv} }
\end{figure*}

The five panels reveal that for each type of source (whatever the light-generation mechanism, optical spectra, radiation pattern, and coherence properties) the photon-count fluctuations exhibit a $1/f^2$ spectral signature over the frequency range $1.0 \times 10^{-6} \leqslant f \leqslant 5.0 \times 10^{-4}$~Hz, corresponding to $33$~min $ \leqslant T_f \leqslant 11.6$~d, where $T_f \equiv 1/f$. The lower frequency limit is governed  by the duration of the individual experiments while the upper frequency limit is established by the photodetector noise floor. 

Averaging the 23 periodograms displayed in Figs.~\ref{bothphotonpg}(a)--\ref{bothphotonpg}(e) that are constructed from data sets of sufficient length yields the black curve labeled ``Mean'' in Fig.~\ref{bothphotonpg}(f). The averaging process is described in Sec.~III~A of~\cite{mohan20PRR}. The averaging leads to improved statistical accuracy, with approximately $2\frac12$ log-units of $1/f^2$ behavior (dashed line).
The small hump in the curve at $1.16 \times 10^{-5}$~Hz, designated by a vertical red line, corresponds to diurnal variations (1~d = 86\,400~s). 

\subsection{Photon-Count Scalograms}\label{Results:PCSG}

Photon-count scalograms for the five classes of optical sources investigated are presented in Fig.~\ref{bothphotonwv}. As with the photon-count periodograms portrayed in Fig.~\ref{bothphotonpg}, the data were collected using the experimental configuration displayed in Fig.~\ref{setup}(a). The scalograms were computed using Daubechies 4-tap wavelets~\cite{daubechies92} via the procedure described in Sec.~II~F of~\cite{mohan20PRR}.
Figures~\ref{bothphotonwv}(a)--(e) represent collections of individual scalograms for LEDs, SLEDs, LDs, incandescent sources, and a betaluminescent source, respectively. 

In analogy with the periodograms, the scalogram for each individual source, whatever its light-generation mechanism, optical spectrum, radiation pattern, and coherence properties, exhibits a $T^2$ scaling signature. 

\begin{figure*}[tb!]
	\includegraphics[width=14cm]{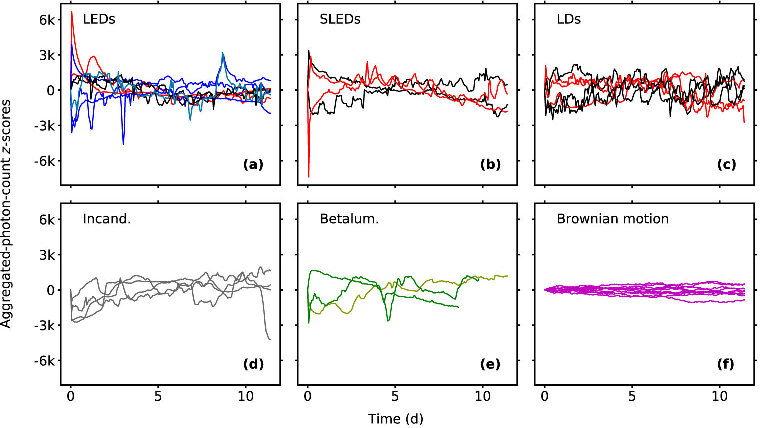}
	\caption{Panels (a)--(e) display aggregated-photon-count $z$-scores for individual sources, plotted one atop the other, for the five classes of optical sources. The initial value of each curve was set to zero to facilitate comparison. The colors in which the curves are plotted correspond to the colors of the light emitted by the individual devices (black and gray represent near-infrared and white incandescent light, respectively; magenta represents simulated Brownian motion). The data  presented were collected using the SPAD module, except where indicated. (a)~LEDs (eight data sets) --- the cyan curve represents data collected with the PMT photodetection system; (b)~SLEDs (four data sets); (c)~LDs (six data sets); (d)~incandescent sources (four data sets); and (e)~a betaluminescent source  (three data sets) --- the yellow-green curve represents data collected with the PMT photodetection system. All of the curves in this figure represent data sets of the same length (11.6~d) with the exception of the two green curves in panel~(e), whose durations are 8.7~d and 9.3~d. Although measured at different times and under different laboratory conditions, all of the traces exhibited irregular long-timescale fluctuations, whatever the type of source or its spectrum. (f) Ten simulated sample paths of ordinary 1D Brownian motion (magenta) whose variances are commensurate with the data. (Adapted from \cite[Fig.~6]{mohan20PRR}.)
		\label{brown}}
\end{figure*}

Again, the 23 individual scalograms shown in Figs.~\ref{bothphotonwv}(a)--(e) that are constructed from data sets of sufficient length follow each other sufficiently closely that they may be averaged to improve statistical accuracy. The outcome of the averaging process, displayed in Fig.~\ref{bothphotonwv}(f), 
increases approximately as $T^2$ (dashed line) over the range $1.0 \times 10^3 \leqslant T \leqslant 1.0 \times 10^5$~s.
The observed $T^2$ growth is a mirror image of the square-law decrease of the observed averaged periodogram portrayed in Fig.~\ref{bothphotonpg}(f). Diurnal variations are again indicated by a vertical red line (1~d = $8.64 \times 10^4$~s). 
The consistency of the scalograms and periodograms across many runs and sources indicates that the computations for both measures are sound.

\subsection{Photon-Count \boldmath$z$-Scores} \label{Exp:zscores}

Since an inverse-square-law spectrum at low frequencies often signals the presence of an underlying source of Brownian motion
\cite[Sec.~2.4.2]{lowen05}, it is useful to examine the time course of the aggregated photon count to ascertain if that is the case here.
The $z$-score, a measure of the deviation of a random variable from the mean, specified in units of standard deviation, is useful for this purpose. As examples, a $z$-score of 0 indicates that the value of the random variable is the same as that of the mean while a $z$-score of $-2$ indicates that the value of the random variable lies 2 standard deviations below the mean.

Individual aggregated-photon-count $z$-scores for the five classes of optical sources considered are displayed in Figs.~\ref{brown}(a)--\ref{brown}(e). 
These time traces were calculated by summing disjoint sets of 1000 consecutive data values to yield a single aggregated number that corresponds to 5000~s (83.33~min) of photon arrivals for each set. For each curve, the first aggregated value was subtracted from all aggregated values for that curve, with the purpose of yielding a curve whose initial value is zero. 

Close examination of the curves shown in  Figs.~\ref{brown}(a)--(e) reveals that the characteristics of the aggregated-photon-count $z$-scores vary somewhat from one source class to another, as discussed in Sec.~V of \cite{mohan20PRR}. Nevertheless, all of the data that lead to these curves exhibit an inverse-square spectrum at low frequencies and a concomitant quadratic wavelet variance at large scales~\cite{mohan20PRR}. 

For purposes of comparison, the ten magenta curves displayed in Fig.~\ref{brown}(f) portray simulated sample paths for ordinary 1D Brownian motion with commensurate values of the variance. Comparison of the photon-count traces in Figs.~\ref{brown}(a)--\ref{brown}(e) with those in Fig.~\ref{brown}(f) reveals that the photon-count traces are considerably more irregular than those for ordinary Brownian motion.

\section{Low-Frequency Electrical Fluctuations in Electroluminescent and Incandescent Sources} \label{sec:ledsledldincand}

In seeking to establish the origin of the low-frequency fluctuations associated with the photon counts in  electroluminescent and incandescent sources, it is sensible to begin by investigating the spectra and correlations that involve the currents and voltages in the sources themselves --- these measures are examined in Secs.~\ref{ssec:ivperiodograms} and \ref{ssec:cc}, respectively. 

\subsection{Current and Voltage Periodograms} \label{ssec:ivperiodograms}

\begin{figure}[htbp!]
	\centering
	\includegraphics[width=8cm]{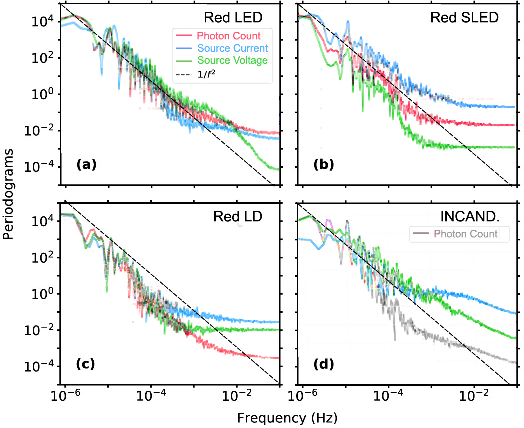}
	\caption{Source current and voltage periodograms, together with concomitantly collected photon-count periodograms, for four types of optical sources operated under constant-current conditions, plotted on doubly logarithmic coordinates: (a)~a red LED (filename 20110423); (b)~a red SLED (filename 20110831); (c)~a red LD (filename 20101029); and (d)~a red incandescent source (filename 20101005). Each panel displays normalized periodograms for the photon count (red or gray), source current (blue), and source voltage (green). The data were collected over a period of 11.6~d using the experimental arrangement depicted in Fig.~\ref{setup}(a) with a SPAD module. All of the periodograms decrease roughly as $1/f^2$ (dashed lines) over the frequency range $1.0 \times 10^{-6} \leqslant f \leqslant 5.0 \times 10^{-4}$~Hz, despite the fact that they represent different variables. Aside from making use of different types of sources, the four experiments were carried out at different times and under different laboratory conditions. \label{pcv} } 
\end{figure}

We augment the photon-count periodograms portrayed in Figs.~\ref{bothphotonpg}(a)--(d) by juxtaposing them with periodograms for the source currents and voltages. The comparisons are provided in Figs.~\ref{pcv}(a)--(d) for a red LED, a red SLED, a red LD, and an incandescent source, respectively, all operated under constant-current conditions. The light emitted by the incandescent source is actually also red, rather than white, by virtue of the low current (and hence low temperature) at which its filament was operated~\cite[Sec.~4.8]{teich2025}. 

The current and voltage were sampled at a rate of 1~s$^{-1}$ with the help of a Keithley Instruments power supply (SourceMeter 2400).
Since the current and voltage periodograms displayed in Fig.~\ref{pcv} closely resemble those of the photon counts, it is sensible to wonder whether low-frequency current and/or voltage fluctuations may underlie the low-frequency photon-number fluctuations. This rationale turns out not to be  plausible, however, since the results for betaluminescence reported in Sec.~\ref{sec:betalum} reveal that an applied electrical drive is not even necessary for generating low-frequency photon-count fluctuations. Nevertheless, the current and voltage may serve as proxies, mediators, or diagnostics in specific  operating modes.
	
\subsection{Correlation Coefficients}\label{ssec:cc}

To further examine the relationships among the photon counts, device currents, and device voltages, we calculate the dependencies of various correlation coefficients that involve these variables on the size of the counting window $T$, over the range $1 \leqslant T \leqslant 10^3$~s.
A preliminary analysis of this nature was reported in~\cite[Chap.~4]{mohan10}.

The experiments and results considered in this section differ from those reported in the other sections of this paper in the following ways: 1)~the source was a single red AlInGaP LED with center wavelength $\lambda = 630$~nm (Kingbright WP1513SURC/E); 2)~the experimental arrangement was similar to that portrayed in Fig.~\ref{setup}(b), but made use of an identical pair of SPAD modules~\cite[Fig.~4.1, Perkin–Elmer SPCMARQ-15-FC]{mohan10} rather than a SPAD and a PMT; 3)~the data  were collected over a period of $\approx 10~\text{h} = 3.6 \times 10^4$~s, rather than over $11.6~\text{d} \approx 1.0 \times 10^6$~s, which limits the counting window to $T \leqslant 10^3$~s to ensure sufficient statistical accuracy; 4)~experiments were carried out under both constant-current and constant-voltage conditions; 5)~experiments used two different values of the detected photon-flux ($\approx$~400,000~s$^{-1}$ and $\approx$~100,000~s$^{-1}$); and 6)~the minimum counting window was $T = 1$~s rather than $5$~s.

\begin{figure*}[tb!]
\centering
\includegraphics[width=10.5cm]{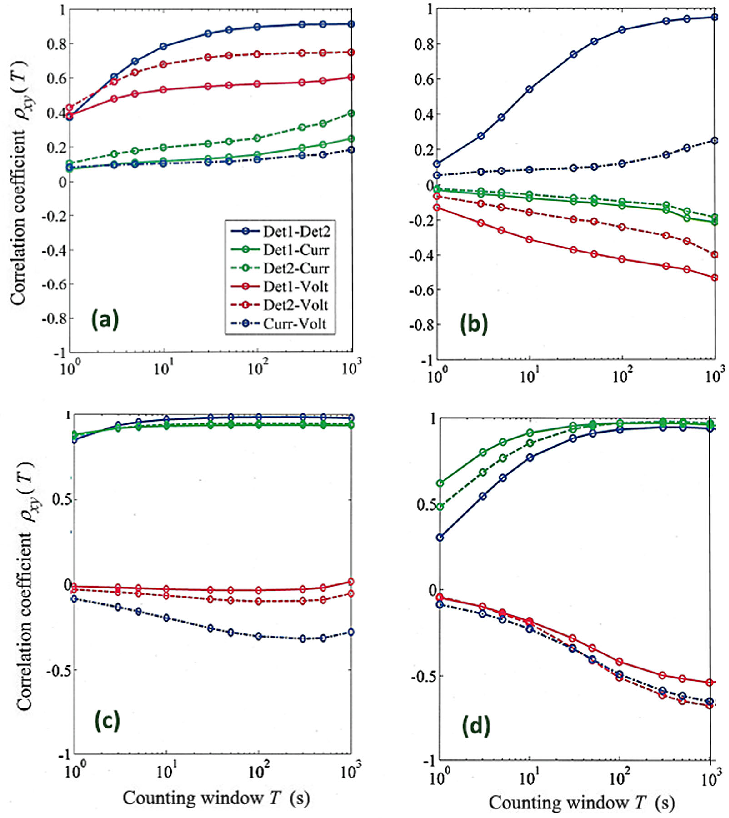}
\caption{Correlation coefficients $\rho_{xy}(T)$ vs. counting windows $T$ for a red LED operated under constant-current and constant-voltage conditions, and measured with a pair of SPADs using an experimental arrangement analogous to that depicted in Fig.~\ref{setup}(b) (filename 20091023). Experiments were carried out both with and without optical attenuation of the LED light (the associated count rates were $\approx 100\,000~\text{s}^{-1}$ and $400\,000~\text{s}^{-1}$, respectively). (a)~Light from the red LED operated at constant current. The key in this panel identifies the curves in all four quadrants of the figure.
(b)~Attenuated light from the red LED operated at constant current. (c)~Light from the red LED operated at constant voltage. (d)~Attenuated light from the red LED operated at constant voltage. 
In all cases, the detector--detector correlations increase with the duration of the counting window. (Panels (c) and (d) adapted from \cite[Fig.~4.18]{mohan10}.)  }
\label{CorrCoeffs}
\end{figure*}

The correlation coefficients
$\rho_{xy}(T)$ for the variables $x$ and $y$, as a function of the counting time $T$, may be expressed as
\begin{equation}
	\rho_{xy}(T)=
	\frac{\sum_{m=1}^{M(T)}(x_m-\overline{x})(y_m-\overline{y})} {
	\sqrt{\sum_{m=1}^{M(T)} (x_m-\overline{x})^2} \:
		\sqrt{\sum_{m=1}^{M(T)} (y_m-\overline{y})^2}
	} \,, 
	\label{eq:corrcoeff}
\end{equation}
where $\overline{x}$ and $\overline{y}$ are the mean values of the $M(T)$ binned samples. 
These correlation coefficients are displayed in Fig.~\ref{CorrCoeffs} for the red LED light using various combinations of the four variables. 

The detector--detector correlation coefficients in all four panels of Fig.~\ref{CorrCoeffs} are seen to grow with increasing counting window, under both constant-current and constant-voltage operation.
At short $T$, Poisson fluctuations and detector-specific noise dominate both photon-count records, which limits their correlation. As $T$ increases, these fast independent contributions are averaged while the slow common component remains. 

Correlation-coefficient plots for blue and infrared LEDs are comparable to those for the red LED shown in Fig.~\ref{CorrCoeffs}, under both constant-current and constant-voltage operation. In all cases, the detector--detector correlations decrease when the LED photon flux is reduced, as expected. An increase in the detector--detector correlations with increasing counting window is also seen for superluminescent diodes and laser diodes operated at constant current, although the growth is less pronounced than that for LEDs. 

The increase of $\rho_{\scriptscriptstyle \rm {DET1,DET2}}(T)$  points to a slowly varying intensity that is shared by the two optical branches. This behavior is consistent with: 1)~the inverse-square baseband spectra illustrated in Fig.~\ref{bothphotonpg}; \,2)~the SPAD/PMT joint-histogram results discussed in Secs.~\ref{sec:detectnoise} and \ref{ssec:statfluctjoint};  and \,3)~the DSPP interpretation developed in Sec.~\ref{ssec:DSPP}.

The current and voltage correlations shown in Fig.~\ref{CorrCoeffs} display a complementary constraint: their connection to the photon count depends on the mode of operation. Under constant-current drive, voltage is the more visible proxy, while under constant-voltage drive, current assumes that role. Thus current and voltage can track the slow photon-rate fluctuations in particular operating modes, but they do not constitute a unique or source-independent cause. This conclusion is reinforced by the persistence of similar low-frequency photon-count behavior under attenuation and, more decisively, by the betaluminescence results presented in Sec.~\ref{sec:betalum}, where no applied current or voltage is involved.

%The statistics of all quantities (mean, CV = stdev/mean, and correlation coefficient) are provided in the file: STATS-102309-ATTN.txt. The data itself is placed in the file: LED-RED-CC-ATTN-102309-1s-65000-D1-D2-C-V.txt. The data file has 4 columns representing, from left to right: Photon counts from detector 1, Photon counts from detector 2, current samples, and voltage samples.
%The folder Figures-102309 contains figures in three formats (.fig [Matlab], .eps, and .jpg) for each quantity. There are five pairs of figures: periodgrams for photon-counts_det1; periodgrams for photon-counts_det2; current; voltage; Zscores. Plotted along with the periodgrams are appropriate parametric fits. 

\section{Low-Frequency Photon-Count Fluctuations in Betaluminescence}\label{sec:betalum}

\begin{figure*}[tb!]
	\centering
	\includegraphics[width=10.7cm]{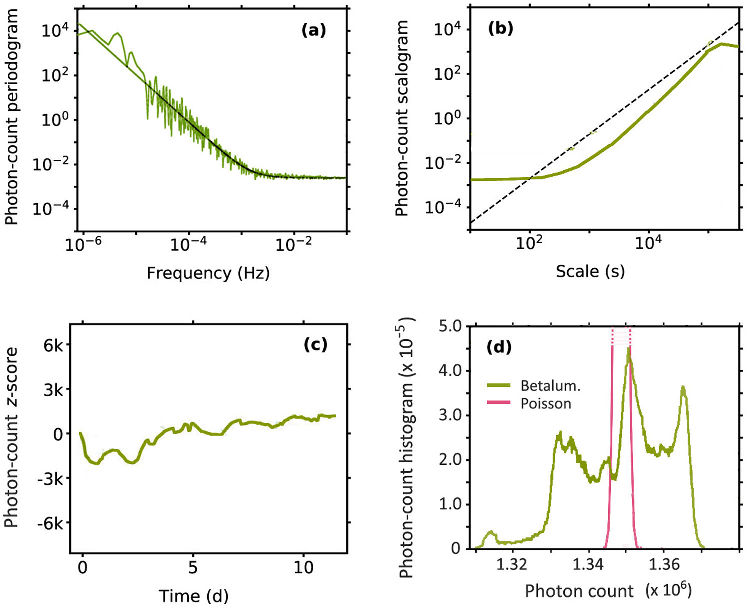}
	\caption{Statistical properties of the photon count for tritium-initiated betaluminescence from a green phosphor. These measures were extracted from photon-count data that were continuously collected for a period of 11.6~d using the PMT photodetection system with a counting window of $T=5$~s (filename 20111127). (a)~The normalized photon-count periodogram, $S_P(f)$ vs. frequency $f$, decreases approximately as $1/f^2$ (solid curve) over the frequency range $1.0 \times 10^{-6} \leqslant f \leqslant 1.0 \times 10^{-3}$~Hz. (b)~The normalized Daubechies 4-tap wavelet variance (scalogram), $A_P(T)$ vs. scale $T$, increases approximately as $T^2$ (solid curve) over the range $1.0 \times 10^3 \leqslant T \leqslant 1.0 \times 10^5$~s, in correspondence 	with the $1/f^2$ decrease of the periodogram. (c)~Aggregated-photon-count $z$-score, with the initial value set to zero; the highly irregular long-timescale fluctuations resemble those associated with electroluminescence and incandescence [Figs.~\ref{brown}(a)--(d)], and differ substantially from the far smoother variations of Brownian motion with commensurate variance  [Fig.~\ref{brown}(f)]. (d)~Betaluminescence photon-count histogram. The mean count is $\overline{n}_{\scriptscriptstyle \rm PMT} \approx 1.35 \times 10^6$, resulting from a count rate $\approx 200\,000$ and a counting window $T = 5$~s. The count variance is clearly far greater than that for a Poisson photon-counting distribution of the same mean (red curve) and the distribution is more irregular. \label{H3statistics} }
\end{figure*}

The photon-count fluctuations associated with betaluminescence offers a number of key insights beyond those gleaned from the analyses of electroluminescence and incandescence provided in Sec.~\ref{sec:ledsledldincand}.
We begin by reviewing the statistical properties of betaluminescence (Sec.~\ref{betalumlfspectsignif}) and continue by highlighting the unique perspective that this form of light provides (Sec.~\ref{betalumimportance}).

\subsection{Statistical Properties of Betaluminescence}\label{betalumlfspectsignif}

The statistical properties of the photon count for tritium-initiated betaluminescence from a green phosphor are highlighted in Fig.~\ref{H3statistics}.

In common with the curves for the electroluminescent and incandescent optical sources displayed in Figs.~\ref{bothphotonpg}--\ref{brown}, the periodogram portrayed in Fig.~\ref{H3statistics}(a) exhibits $1/f^2$ low-frequency spectral behavior, the scalogram shown in Fig.~\ref{H3statistics}(b) increases as $T^2$, and the aggregated-photon-count $z$-score presented in Fig.~\ref{H3statistics}(c) meanders in an irregular fashion that is distinctly different from the paths followed by ordinary Brownian motion [Fig.~\ref{brown}(f)]. 

Furthermore, the photon-count histogram plotted in Fig.~\ref{H3statistics}(d) is far broader and more irregular than the Poisson counting distribution of the same mean (displayed in red). The large variance is not a signature of detector-noise alone; rather, it represents precisely the kind of overdispersion expected when otherwise Poisson photon counts are mixed by a slowly varying random intensity function to form a doubly stochastic photon-counting distribution, as described in Sec.~\ref{ssec:DSPP}. 

\subsection{Insights from Betaluminescence}\label{betalumimportance}

The photon-count fluctuations associated with betaluminescence provide a unique perspective in connection with our search for the origin of the low-frequency photon-count spectra. That is because the betaluminescence is initiated by a statistically stationary $\beta$-decay process in $^3$H (tritium), rather than by an externally applied current or voltage. The half-life of this isotope is $t_{1/2} \approx 12.32$~yr, so the mean rate is essentially constant over the duration of the experiment.

In short, betaluminescence has a number of unique features that are useful for unraveling the underlying origin of the source-independent photon-count fluctuations at baseband:
\begin{itemize}
	\item It is not generated by an applied current or voltage.
	\item It does not involve the consumption of electrical power.
	\item Its excitation is initiated by nuclear decay, whose mean rate is essentially constant. 
\end{itemize}

These features all point away from the physical functioning of an optical source as being responsible for the presence of low-frequency photon-count fluctuations. And they support the notion that the corresponding low-frequency photon-count fluctuations in electroluminescence and incandescence do not originate from current and/or voltage fluctuations internal to the device, as discussed in Sec.~\ref{ssec:ivperiodograms}.

\section{Low-Frequency Source-Temperature Fluctuations}
\label{Results:STPG}

\begin{figure}[htbp!]
	\centering
	\includegraphics[width=7cm]{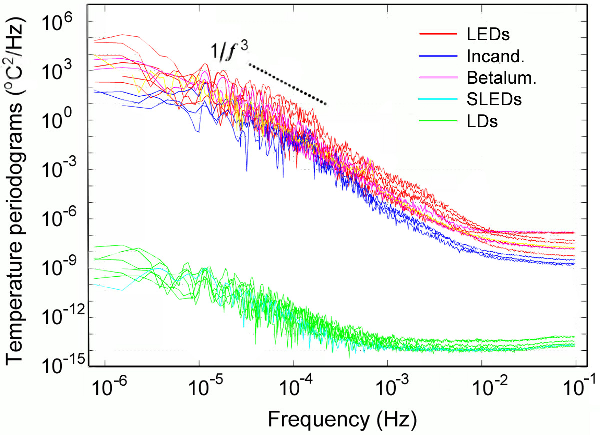}
	\caption{Unnormalized source-temperature periodograms collected during experiments with various classes of light sources. The upper cluster of curves, plotted one atop the other, are periodograms, $S_\mathcal{T}(f)$ vs. $f$, recorded during individual experiments with LEDs, incandescent sources, and a betaluminescent source, all of which operate at ambient temperature. The lower cluster of curves portray  periodograms recorded during individual experiments with SLEDs and LDs, which required cooling, again plotted one atop the other. In the region where both sets of curves are decreasing, the lower cluster lies some 10--12 orders of magnitude below the upper cluster, indicating that source-temperature fluctuations are not responsible for the baseband behavior of the photon-count periodograms.
	Moreover, the temperature records differ from the photon-count records in spectral form.}
	\label{temppgs}
\end{figure}

We now proceed to confirm that source-temperature fluctuations are not responsible for the $1/f^2$ form of the photon-count periodograms displayed in Figs.~\ref{bothphotonpg} and \ref{H3statistics}(a). In Fig.~\ref{temppgs} we present a collection of \textit{unnormalized} source-temperature periodograms, collected using a Thorlabs temperature sensor, that were concomitantly recorded with the photon-count periodograms.

The upper cluster of curves are temperature periodograms for experiments carried out with devices that did not require cooling (LEDs, incandescent sources, and the betaluminescent source), so that the source temperature was the same as the recorded ambient temperature.
The best fit to this cluster of curves over the frequency range $1.0 \times 10^{-5} \leqslant f \leqslant 5.0 \times 10^{-3}$~Hz is a straight line representing behavior of the form $1/f^3$, in accord with other observations~\cite{lowen20}.

The lower cluster of curves are temperature periodograms for experiments carried out with devices that required cooling (SLEDs and LDs), so that the temperature recorded was the local temperature at the source. This cluster of periodograms lies some 10--12 orders of magnitude below the upper (uncooled) cluster in the region where both sets of curves are decreasing.

Since the temperature fluctuations for the upper cluster of curves are a trillion times stronger than those for the lower cluster, and since the temperature periodograms in the upper cluster do not project the $1/f^2$ behavior that is the hallmark of the photon-count periodograms, we conclude that source-temperature fluctuations are not responsible for the low-frequency behavior of the photon-count periodograms.

\section{Low-Frequency Detector Noise and Joint SPAD/PMT Photon-Count Histogram}\label{sec:detectnoise}

It was established above that neither electrical nor temperature fluctuations at the source are responsible for the inverse-square photon-count fluctuations at baseband. In an attempt to further constrain the origin of these fluctuations, we turn now to investigating the noise and joint photon-counting statistics at the detectors.

\begin{figure*}[tb!]
	\centering
	\includegraphics[width=15cm]{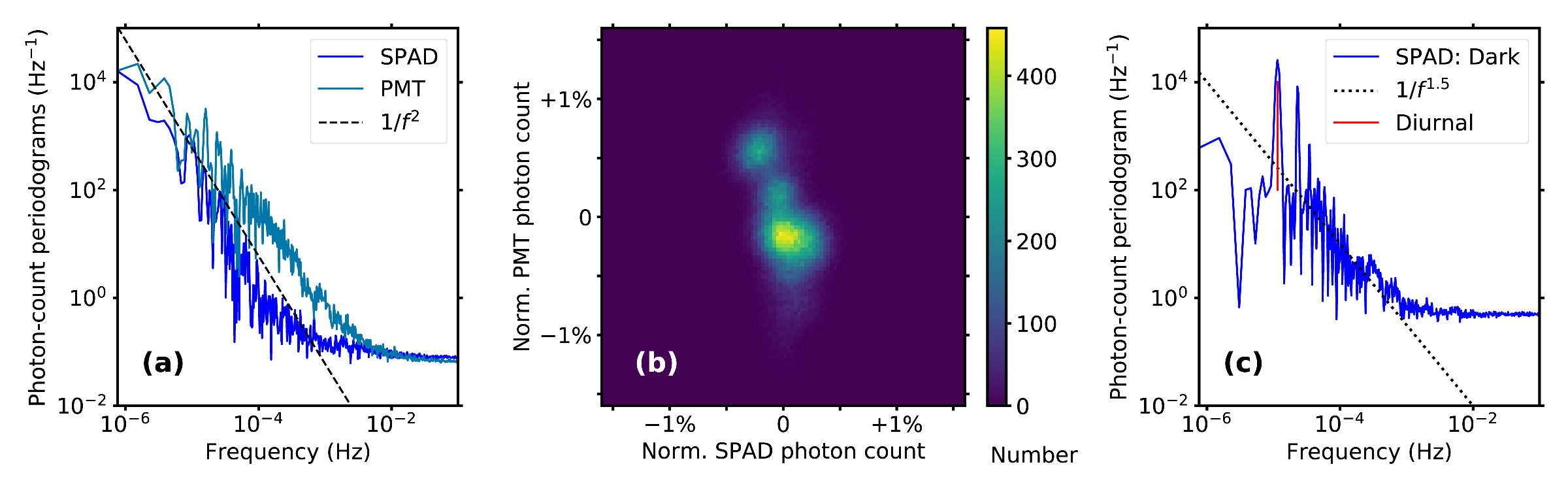}
	\caption{Photon-counting measurements obtained from an experiment using two independent photodetection systems, configured as in Fig.~\ref{setup}(b). (a)~Normalized photon-count periodograms, $S_P(f)$ vs. frequency $f$, obtained using the SPAD module and the PMT photodetection system, plotted on doubly logarithmic coordinates. These curves, which are among the eight LED curves presented in Fig.~\ref{bothphotonpg}(a), are drawn from an experiment of duration 11.6~d using a blue LED (filenames 20111214a and 20111214b for the SPAD and PMT, respectively).
	Both curves exhibit approximate $1/f^2$ spectral behavior (dashed line); close examination reveals that the details of their fluctuations are substantially similar. The baseline noise of the SPAD module sets in at a lower frequency than that of the PMT system, however. (b)~Joint photon-count histogram displaying the photon counts registered by the SPAD (abscissa) and the PMT (ordinate),  normalized to their respective values of the mean count, in a sequence of consecutive counting windows of duration $T = 5$~s. The joint histogram, constructed from a 7.51-d subset of the 11.6-d data set, exhibits positive overall correlation and a narrow dominant peak, indicating that the two detectors register similar results.
	(c)~Normalized photon-count periodogram for the SPAD module in the absence of deliberate illumination, plotted on doubly logarithmic coordinates.
	This curve is drawn from an experiment of duration 11.6~d (filename 20101227). Over the central frequency region, the spectrum behaves approximately as $1/f^{3/2}$ (dotted curve), which differs from the spectra that emerge in the presence of illumination, as displayed in panel~(a) of this figure and in
	Figs.~\ref{bothphotonpg} and \ref{H3statistics}(a).
	The peak in the curve at $1.16 \times 10^{-5}$~Hz, indicated by the vertical red line, and other peaks at harmonics thereof, correspond to diurnal variations (1~d = 86\,400~s), a consequence of the residual leakage of a small amount of daylight into the laboratory.
	(Adapted from \cite[Fig.~3]{mohan20PRR}.)}
	\label{APD+PMT}
\end{figure*}

We demonstrate in the following that two independent photon-counting systems, one based on the SPAD module and the other based on the PMT system, yield similar results even though the two photodetection systems have very different properties and modes of operation. It will become apparent that the $1/f^2$ behavior of the photon-count periodograms presented in Figs.~\ref{bothphotonpg}, \ref{H3statistics}(a), and \ref{APD+PMT}(a) does not arise from noise or temperature fluctuations at the detectors.

Using the experimental configuration depicted in Fig.~\ref{setup}(b), in which data are simultaneously acquired by a SPAD and a PMT, we obtain the normalized photon-count periodograms displayed in Fig.~\ref{APD+PMT}(a). The details of the two curves are seen to be quite similar --- and both exhibit approximate inverse-square spectral behavior (dashed line). The principal distinction between the two periodograms is the frequencies at which they reach baseline. The baseline noise for the SPAD becomes appreciable at $f \approx 5 \times 10^{-4}$~Hz, whereas the PMT is capable of registering fluctuations over an additional log-unit of frequency. 

The joint photon-count histogram for the two photodetectors, plotted on axes whose values are normalized by their respective mean photon counts, is displayed in Fig.~\ref{APD+PMT}(b). The data exhibit a narrow dominant peak, with a width that lies well within 1\% of the normalized photon-count values, and have positive overall correlation. The numbers of observed events are indicated by the color key just to the right of panel~(b). With a counting time $T = 5$~s, and an experiment of duration 7.51~d, the means and standard deviations of the photon counts, using the SPAD and PMT photodetection systems  respectively, were  
 $\overline{n}_{\scriptscriptstyle \rm SPAD} = 985\,370$, $\sigma_{\scriptscriptstyle \rm SPAD} = 1\,805$; and $\overline{n}_{\scriptscriptstyle \rm PMT} = 995\,441$,  
$\sigma_{\scriptscriptstyle \rm PMT} = 3\,942$. 
The mean photon counts differ slightly because the detectors were not presented with the same values of the photon flux and they have different quantum efficiencies --- the plot in Fig.~\ref{APD+PMT}(b) thus relies on axes that are normalized to their mean values. 

It is apparent from Figs.~\ref{APD+PMT}(a) and \ref{APD+PMT}(b) that the results obtained using the two photodetectors are  reasonably similar, confirming that our measurement methods are sound and that our results are consistent. The SPAD and PMT systems operate on the basis of different principles, and exhibit different noise properties, temperature dependencies, and quantum efficiencies. The two detectors also make use of largely independent timing, counting, and software chains. 
The results presented in Fig.~\ref{APD+PMT}(a) thus indicate that the $1/f^2$ behavior of the photon-count periodograms does not appear to originate from noise or temperature fluctuations at the detector. 

We show in Fig.~\ref{APD+PMT}(c) that in the absence of light, the SPAD module exhibits a spectrum that falls off as $\approx 1/f^{3/2}$ at these low frequencies, rather than as $\approx 1/f^2$, as it does in the presence of light.
The contrast between illuminated and dark SPAD operation thus indicates that the inverse-square spectral fluctuations are tied to the detected optical flux, rather than to intrinsic free-running detector fluctuations.
Moreover, the aggregated-count $z$-score fluctuations are far smaller in the dark than in the presence of light.
Lastly, it should be mentioned that PMT dark counts can be distinctly non-Poisson and quite clustered as a result of contributions from cosmic-ray showers, alathough this contribution has a small mean~\cite{campos1987cosmicray}.

\section{Low-Frequency Technical Noise}\label{sec:technoise}

Given that the origin of the $1/f^2$ photon-count spectra at baseband cannot be ascribed to fluctuations of the current, voltage, power, or temperature of the source, nor to noise or temperature fluctuations at the detector,
we proceed to examine possible technical sources of noise.

These fall into three categories, which are considered in turn: A)~optical-system noise; B)~local environmental noise and mechanical perturbations; and C)~source package noise. It will become clear that none of these is a plausible  source of the $1/f^2$ photon-count spectra. 

\subsection{Optical-System Noise}\label{ssec:technoiseoptics}

Optical-system noise, including optical-throughput fluctuations, beam-pointing fluctuations, and aperture effects, are unlikely to be sources of the baseband noise for the following reasons:
\begin{itemize}
	\item \(1/f^2\) behavior persists whatever the attenuation of the beam.
	\item The SPAD and PMT detectors have relatively large active areas, making pointing and aperture limitations unlikely to dominate.
	\item It is implausible that optical-throughput effects, random beam-pointing, or aperture effects would repeatedly yield the same \(1/f^2\) spectral law across many source classes and experimental runs.
\end{itemize}

\subsection{Local Environmental Noise and Mechanical Perturbations}\label{ssec:technoiseair}

Local environmental noise, such as air-index and laboratory-air fluctuations, as well as mechanical perturbations of the experimental apparatus, are also not likely to be sources of the baseband noise for the following reasons:
\begin{itemize}
	\item The experiments, carried out over a period of approximately 1.5 years, all yielded the same inverse-square baseband spectrum.
	\item Consistent results were obtained across seasons, as well as across large variations in air-index and laboratory-air conditions. 
	\item The \(1/f^2\) behavior was the same whether climate control in the laboratory was active or suspended.
	\item The same behavior was observed under different thermal conditions, indicating that effects such as the thermal expansion and contraction of optical components were unimportant.
	\item Many distinct mechanical configurations, sources, source packages, and detectors would be highly unlikely to produce the same inverse-square-law behavior for all experimental runs.
	\item The effects of microscopic seismic vibrations transmitted through the optical bench should be unimportant in intensity-based (noninterferometric) systems. 
\end{itemize}

\subsection{Source Package Noise}\label{ssec:technoisesourcepackage}

It is implausible that source-package effects could be the underlying cause of the low-frequency photon-number fluctuations. The many sources used involve different emission mechanisms and distinct package types, and all yield approximately the same \(1/f^2\) spectrum.

\section{Summary of Excluded Sources of the Baseband Photon-Count Fluctuations}\label{sec:technoiseexclude}

Based on the considerations detailed in Secs.~\ref{sec:ledsledldincand}--\ref{sec:technoise}, the presence of the $1/f^2$ low-frequency photon-count fluctuations cannot plausibly be ascribed to any of the following:
\begin{itemize}
	\item Current/voltage/power fluctuations at the source.
	\item Temperature fluctuations at the source.
	\item Semiconductor-junction phenomena.
	\item Noise fluctuations at the detector.
	\item Temperature fluctuations at the detector.
	\item Optical-system noise.
	\item Local environmental noise.
	\item Mechanical perturbations of the experiment.
	\item Source package noise.
\end{itemize}

Based on the information distilled to this point, the remainder of this paper is devoted to characterizing the statistical nature of the observed low-frequency fluctuations and to exploring a number of possible heterodox hidden origins of the enigmatic $1/f^2$ photon-count spectra at baseband.

\section{Statistical Characteristics of the Observed Low-Frequency Photon-Count Fluctuations}\label{sec:statfluct}

The central empirical observation reported in Secs.~\ref{Exp} and \ref{sec:betalum} is that a broad variety of optical sources exhibit irregular, long-duration photon-count fluctuations with an approximately inverse-square baseband spectrum,
\begin{equation}
	S(f) \propto \frac{1}{f^2}\,.
\end{equation}
This behavior is present over the approximate frequency range
\begin{equation}
	10^{-6}\ \mathrm{Hz} \leqslant f \leqslant 5\times 10^{-4}\ \mathrm{Hz},
\end{equation}
corresponding to timescales ranging from tens of minutes to 11.6~d (the duration of the individual experiments).

We seek to elucidate the origin of this spectral behavior.
Further information can be gleaned from the spectra of the original and shuffled versions of the photon counts, which are compared in Sec.~\ref{sec:statfluctrate}; and from the time evolution of the joint SPAD/PMT photon-count histogram, which is presented in Sec.~\ref{ssec:statfluctjoint}.  
Both observations suggest that the observed fluctuations arise from a slow stochastic modulation of the photon-count intensity. A plausible and widely used model for characterizing behavior of this kind is based on the doubly stochastic Poisson point and counting processes described in Secs.~\ref{ssec:DSPP} and \ref{ssec:DSPPmath}, respectively.

\subsection{Photon-Count Shuffling}\label{sec:statfluctrate}

Photon-count shuffling is a process that randomizes the  photon-count ordering. This serves to destroy any correlations or dependencies that might have
existed among the photon counts, while leaving the 
photon-count histogram intact.
The shuffling operation proves valuable for determining whether the observed behavior stems from the ordering of the photon counts or from the form of the photon-count histogram (or from both). 
Shuffling is often used in connection with the interevent intervals of a point process, and is but one of a multitude of operations that create surrogate data sets that prove useful for elucidating the underlying nature of point processes~\cite[Chap.~11]{lowen05}.

\begin{figure}[htbp!]
	\centering
	\includegraphics[width=4.75cm]{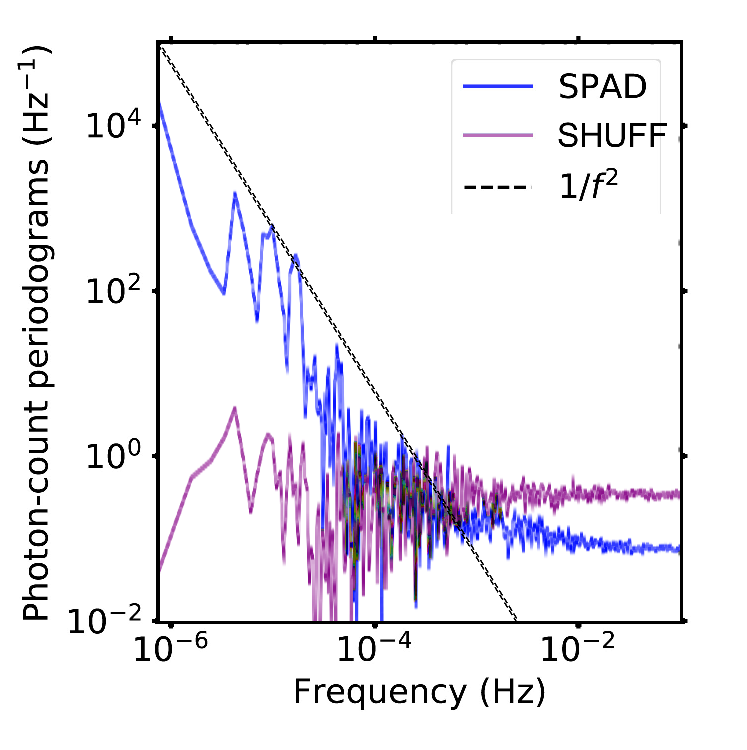}
	\caption{Normalized photon-count periodogram, $S_P(f)$ vs. frequency $f$ (blue curve), plotted on doubly logarithmic coordinates. The periodogram constructed from the shuffled photon counts is represented by the maroon curve. The underlying data for these curves, which are representative, have their origin in an experiment of duration 11.6~d that made use of a blue LED (filename 20111214a) and a SPAD module. The blue curve exhibits approximately $1/f^2$ spectral behavior (dashed line), just as does the blue curve depicted in Fig.~\ref{APD+PMT}(a), which represents the same data set. Shuffling the photon counts of the original data set destroys their ordering and, with it, the signature of the original rate variation of the process; the result is a flat photon-count periodogram.}
	\label{FigShuffle}
\end{figure}

We consider photon-count shuffling in the context of the photon-count periodogram. A normalized photon-count periodogram of representative  data, portrayed as the blue curve in Fig.~\ref{FigShuffle}, exhibits $1/f^2$ spectral behavior. On the other hand, the periodogram of the shuffled photon counts, represented as the maroon curve in Fig.~\ref{FigShuffle}, exhibits no such structure, but rather is consistent with a white-noise (uncorrelated) process of photon counts. 

The flat form of the maroon curve indicates that the $1/f^2$ spectral behavior evident in the blue curve arises from the ordering of the photon counts, signifying the presence of a hidden varying intensity function. 
This interpretation is consistent with the detector–detector correlations reported in Sec.~\ref{ssec:cc}.

\subsection{Dynamical Representation of the Joint Photon-Count Histogram}\label{ssec:statfluctjoint}

Another measure that offers a window onto the origin of the low-frequency spectral behavior is the joint SPAD/PMT photon-count histogram. A static version of that statistic, shown in Fig.~\ref{APD+PMT}(b), displays the photon count registered by the PMT on the ordinate vs. that registered by the SPAD on the abscissa, integrated over a period of 7.51~d and normalized to their respective mean counts. 

\begin{figure}[htbp!]
	\centering
	\includegraphics[width=7.0cm]{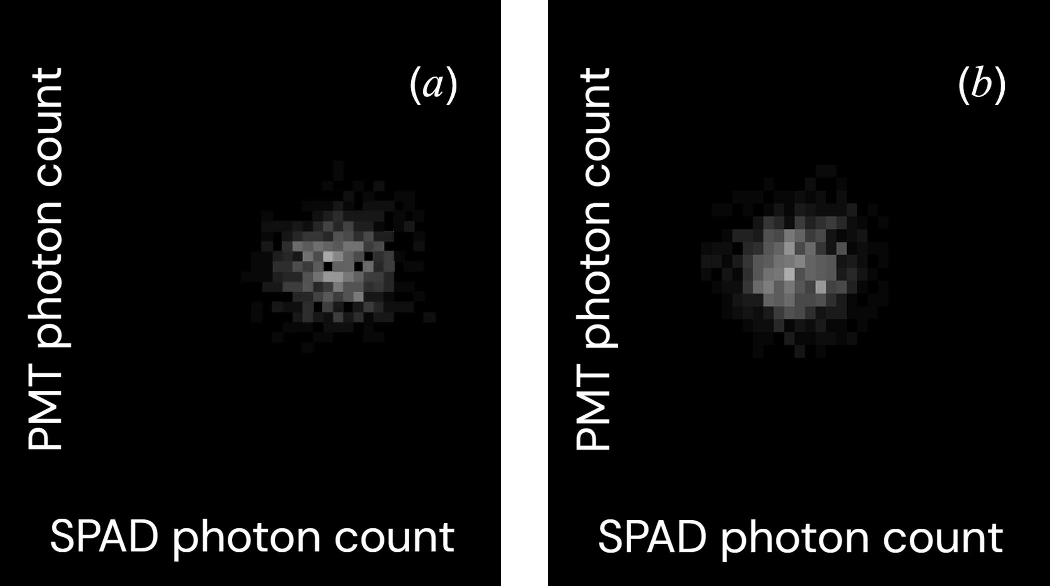}
	\caption{(a)~Representative frame from the animated joint SPAD/PMT photon-count histogram provided in the Ancillary Directory.
	The video displays the PMT photon count (ordinate) vs. the SPAD photon count (abscissa) for light from a blue LED over a period of 7.35~d. This subset of the 11.6-d data sets (filenames 20111214a and 20111214b) is nearly identical to the 7.51-d static subset displayed in Fig.~\ref{APD+PMT}(b).
	Each of the 127 frames of the video comprises 1000~windows, each of 5-s duration, so that each frame 
	represents a duration of $5000$~s $\approx 83.33$~min $\approx 1.39$~h. 
	Since the mean count rate registered in both the PMT and SPAD branches of the experiment [Fig.~\ref{setup}(b)] is $\overline{\mu} \approx 200\,000$~s$^{-1}$, each frame represents an average of $\approx 200\,000 \times 5\,000 = 10^9$~photon counts. The full video is therefore constructed from $\approx 1.3 \times 10^{11}$~photon-count registrations in each detection branch. 
	(b)~Static joint SPAD/PMT photon-count histogram for  simulated Poisson counts with the same mean values as those in (a).}
	\label{PMT-SPAD-simulation}
\end{figure}

The time evolution of the joint histogram, provided as a video in the Ancillary Directory, offers additional insight into the underlying process.
A representative frame from this video is displayed in Fig.~\ref{PMT-SPAD-simulation}(a) and, for purposes of comparison, a static joint histogram for simulated  Poisson counts with the same mean values is provided in Fig.~\ref{PMT-SPAD-simulation}(b).

The video displays a narrow moving peak and positive overall correlation, indicating that the two detectors register similar results.
The peak undergoes mostly diagonal and vertical movements as the video progresses. The diagonal movements indicate that the PMT and SPAD counts vary in concert, 
suggesting the presence of common-mode variability before the two independent detection outcomes. 
The presence of vertical movements reflects excess variability in the PMT output, which is of undetermined origin. The relative absence of horizontal movements indicates that there is little excess variability in the SPAD response. 
(The meandering motion of the central spot mirrors the relative motions of the blue and cyan $z$-score time traces portrayed in Fig.~\ref{brown}(a).)

\subsection{Doubly Stochastic Poisson Point Process}\label{ssec:DSPP}

\begin{figure}[htbp!]
	\includegraphics[width=6.5cm]{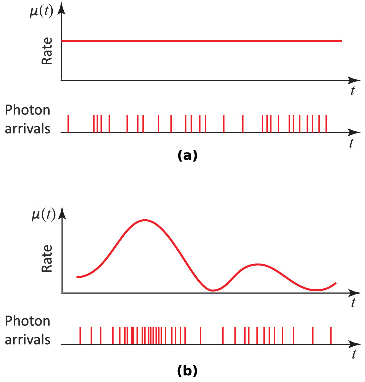}
	\caption{(a)~A fixed-intensity driving process $\,\mu(t) = \mu \,$ and a  sample of the ensuing randomly arriving photons for a homogeneous Poisson point process (HPP). (b)~A time-varying driving process $\mu(t)$ and a sample of the resulting photon arrivals for a DSPP. The density of events tends to be greater for larger values of the intensity $\mu(t)$, although the inherent Poisson randomness limits this association to a probabilistic one.}\label{FigDSPP}
\end{figure}

The inverse-square-law photon-count spectrum at baseband is evidently most directly attributable to fluctuations of the counting-process intensity.
The most economical statistical description for a process of this nature assumes photon registrations that are conditionally Poisson, with a slowly varying intensity. This widely used generalization of the homogeneous Poisson process depicted in Fig.~\ref{FigDSPP}(a) emerges when the intensity $\mu$ varies in time, as illustrated in Fig.~\ref{FigDSPP}(b). The ensuing \textit{doubly stochastic Poisson point process} (DSPP) arises when $\mu(t)$ is taken to be a positive-valued continuous-time stochastic intensity process~\cite[Sec.~4.3]{lowen05}. 

As displayed in Fig.~\ref{FigDSPP}(b), this point process is \emph{doubly random}: one source of randomness arises from the (unobserved) fluctuations of the stochastic intensity $\mu(t)$ and the other is associated with the Poisson event-generation process itself, given the intensity $\mu(t)$. 

By way of illustration, the random intensity variations exhibited by the continuous red curve in Fig.~\ref{FigDSPP}(b) are transformed to (bunched) random photon arrivals.
This process was initially conceived by David Cox in 1955~\cite{cox1955}, and is therefore sometimes called a Cox process. The appellation \textit{doubly stochastic Poisson process} originated with Bartlett in 1963~\cite{bartlett1963}.
The DSPP finds wide use in characterizing many phenomena in the physical and biological sciences.

For completeness, we note that the designation \textit{mixed Poisson process}, introduced in 1955 by Bartlett~\cite{bartlett1955}, signifies that the intensity is a random variable rather than a random process. However, the concept of the DSPP counting distribution stretches back to the studies of Greenwood and Yule in 1920~\cite{greenwood1920}, who studied the Poisson transform of the gamma density, which gives rise to the negative-binomial distribution.

The DSPP is used extensively to model the registration of photons in optics and photonics~\cite{saleh78,cox1980,teich1981a,saleh1982,saleh1983,
teich88,lowen05,saleh19}. When the duration of the counting window $T$ is significantly greater than the coherence time of the source $\tau_c$, and the degeneracy parameter $\delta \equiv \overline{\mu} \tau_c$ is sufficiently small that residual bunching is negligible, fluctuations associated with optical coherence are averaged over the counting window and, given the slowly varying intensity function $\mu(t)$, the photon registrations are well approximated as Poisson~\cite{saleh1982}. 

Since the optical sources discussed in Secs.~\ref{Exp} and \ref{sec:betalum} are characterized by coherence times that lie in the fs to $\mu$s regime~\cite[Table 12.1-2]{saleh19}, and given that $\overline{\mu} \approx 200\,000$~s$^{-1}$ and $T = 5$~s, we have $T/\tau_c \gg 1$ and $\delta=\overline{\mu}\tau_c \lesssim 0.1$ for $\tau_c \lesssim 0.5~\mu$s. Optical-coherence-related bunching is therefore substantially reduced in these sources, and is negligible for sources with shorter effective coherence times. 

It is important to emphasize, however, that even in this case, an external slowly varying intensity function $\mu(t)$, with an effective fluctuation time $\tau_\mu$, 
can nevertheless give rise to a doubly stochastic Poisson point process with an enhanced low-frequency spectrum. We follow this line of reasoning and consider doubly stochastic Poisson counting processes whose intensities exhibit $1/f^2$ spectral properties.

\subsection{Mathematical Properties of the Doubly Stochastic Poisson Counting Process}\label{ssec:DSPPmath}

Formally, for the DSPP we have 
\begin{equation}
	\lim_{\epsilon \to 0} \tfrac{1}{\epsilon} \Pr\left\{N(t + \epsilon) - N(t) > 0 \,
	\big| \, \mu(t)\right\} = \mu(t), \label{eq:dsppdef}
\end{equation}
where $N(t)$ represents the counting process. Two statistics follow immediately from
Eq.~(\ref{eq:dsppdef}). 
Forming the expectations $\E[\cdot]$ of both sides of Eq.~(\ref{eq:dsppdef})
yields 
\begin{equation}
	\E \! \left[\frac{d N(t)}{d t}\right] = \E[\mu],
\end{equation} 
and integration leads to
\begin{equation}
	\E \! \left[N(t)\right] = \E[\mu]\, t.
\end{equation}
For consecutive, nonoverlapping counting windows of duration $T$, the corresponding event count in the $m$th bin is
\begin{equation}
	Z_m = N((m+1)T) - N(mT),
	\label{eq:Zmdef}
\end{equation}
which has mean
\begin{equation}
	\E \! \left[Z_m\right] = \E[\mu]\, T.
	\label{eq:Zmmean}
\end{equation}

In the following expressions involving correlation functions and spectra, we assume that the intensity process is stationary over the analyzed band, or equivalently that the observed $1/f^2$ behavior is regularized by finite-duration or low-frequency cutoffs. Let
\begin{equation}
	\bar{\mu} = \E[\mu],
	\qquad
	\Delta\mu(t) = \mu(t)-\bar{\mu},
\end{equation}
and define the covariance of the intensity fluctuations as
\begin{equation}
	C_\mu(t)=\E\!\left[\Delta\mu(s)\,\Delta\mu(s+t)\right].
	\label{eq:Cmudef}
\end{equation}

Similarly, employing Eq.~(\ref{eq:dsppdef}) at two different times gives rise to the coincidence rate. If
\begin{equation}
	n(t) \equiv \frac{dN(t)}{dt}
\end{equation}
denotes the formal point-process density, then
\begin{eqnarray}
	G_N(t) & \equiv & \E \! \left[n(s)\,n(s+t)\right] \nonumber \\[1mm]
	& = & \E \! \left[\mu(s)\,\mu(s+t)\right] + \bar{\mu}\,\delta(t) \nonumber \\[1mm]
	& = & \bar{\mu}^{\,2} + C_\mu(t) + \bar{\mu}\,\delta(t).
	\label{eq:dsppcrracf}
\end{eqnarray}

By virtue of the Wiener--Khinchin theorem~\cite[Sec~12.1B]{saleh19}, the Fourier transform of Eq.~(\ref{eq:dsppcrracf}) leads to
\begin{equation}
	S_N(f) = \bar{\mu}^{\,2}\delta(f) + S_{\Delta\mu}(f) + \bar{\mu},
	\label{eq:dspppsdfull}
\end{equation}
where $S_N(f)$ denotes the spectrum of the point process associated with $N(t)$, equivalently the spectrum of the formal derivative $d N(t)/d t$, and $S_{\Delta\mu}(f)$ represents the spectrum of the intensity fluctuations $\Delta\mu(t)$. Away from $f=0$, or equivalently for the mean-subtracted point process, Eq.~(\ref{eq:dspppsdfull}) reduces to
\begin{equation}
	S_N^{(c)}(f) = S_{\Delta\mu}(f) + \bar{\mu}.
	\label{eq:dspppsd}
\end{equation}

Apart from the additive Poisson shot-noise floor, the point-process spectrum thus mimics the intensity-fluctuation spectrum  --- in particular, the observation that $S_N^{(c)}(f) \propto 1/f^2$ implies that  $S_{\Delta\mu} (f) \propto 1/f^2$ over the same frequency range, provided the intensity-fluctuation term dominates the white Poisson contribution.

For finite-duration counting windows, the counting operation introduces the rectangular-window transfer function, followed by sampling at intervals $T$. The count in the $m$th nonoverlapping bin may be written as
\begin{equation}
	Z_m = \int_{mT}^{(m+1)T} \frac{d N(s)}{d s}\, d s .
	\label{eq:Zmfilter}
\end{equation}
The rectangular-window transfer function is
\begin{equation}
	H_T(f) = \int_0^T \! e^{-i2\pi f s}\, d s
	= T e^{-i \pi f T} \operatorname{sinc}(fT),
	\label{eq:HT}
\end{equation}
where $\operatorname{sinc}(x)=\sin(\pi x)/(\pi x)$. Thus
\begin{equation}
	|H_T(f)|^2 = T^2 \operatorname{sinc}^2(fT).
	\label{eq:HTmag}
\end{equation}

For nonoverlapping experimentally recorded bins, the measured sequence is the sampled output of this integrate-and-dump operation. Sampling can introduce the usual discrete-time aliasing effects, but in the low-frequency range for which aliasing is negligible and for $fT \ll 1$, the rectangular-window factor is approximately $T^2$, so that the inverse-square spectral dependence is preserved in the count sequence.

Other statistics of the point process also derive from those of the intensity process $\mu(t)$. In particular, for the event count $Z_m$ in the window $[mT,(m+1)T]$, the integrated intensity is defined as
\begin{equation}
	\Lambda_m = \int_{mT}^{(m+1)T} \mu(s)\, d s.
	\label{eq:intrteT}
\end{equation}
The event count and factorial moments are expressed as~\cite{saleh78}
\begin{equation}
	\Pr\{Z_m = n\} = \E \! \left\{\frac{\Lambda_m^n \, \exp \!
		\left[-\Lambda_m\right]}{n!}\right\}
	\label{eq:dspppn}
\end{equation}
and
\begin{equation}
	\E \! \left[ Z_m (Z_m-1)\cdots (Z_m-k+1) \right] = \E \!
	\left[\Lambda_m^k\right],
	\label{eq:dsppfactmom}
\end{equation}
respectively.

For the case at hand, the counting window $T = 5$~s is long in comparison with the coherence time of the source $\tau_c$, and short in comparison with 
the effective fluctuation time $\tau_\mu$ of the slowly varying intensity process over the frequency band of interest, i.e., 
\begin{equation}
	\tau_c \ll T \ll \tau_\mu.
	\label{eq:intrteTtau}
\end{equation}
The integrated intensity can therefore be written as 
\begin{equation}
	\Lambda_m \approx T \mu(mT).
	\label{eq:intrteTshort}
\end{equation}

The corresponding mean and variance of the DSPP counts follow from the laws of total expectation and total variance. Since $Z_m$ is conditionally Poisson with conditional mean $\Lambda_m$, we have
\begin{eqnarray}
	\E[Z_m] & = & \E[\Lambda_m], \label{DSPPmeancnt}\\[1.5mm]
	\operatorname{var}(Z_m) & = & \E[\Lambda_m] + \operatorname{var}(\Lambda_m).\label{DSPPvarcnt}
	\label{eq:dsppmeanvar}
\end{eqnarray}
The DSPP variance thus exceeds the Poisson value by  $\operatorname{var}(\Lambda_m)$, which represents the overdispersion produced by the slow fluctuations of the intensity.

For a stationary, or effectively band-limited, intensity process with covariance $C_\mu(t)$, the second term on the right-hand side of Eq.~(\ref{eq:dsppmeanvar}) may be written as
\begin{equation}
	\operatorname{var}(\Lambda_m) =
	\int_{-T}^{T} (T-|t|)\, C_\mu(t)\, d t.
	\label{eq:lambdavar}
\end{equation}
Under the approximation stated in Eq.~(\ref{eq:intrteTshort}), Eqs.~(\ref{DSPPmeancnt}) and (\ref{DSPPvarcnt}) reduce to
\begin{eqnarray}
	\E[Z_m] & \approx & T\, \bar{\mu}, \nonumber \\[1.5mm]
	\operatorname{var}(Z_m) & \approx & T\, \bar{\mu} + T^2 \operatorname{var}[\mu],
	\label{eq:dsppmeanvarapprox}
\end{eqnarray}
respectively.

Count-based measures are often favored over interval-based ones because the information they provide corresponds to the real time of the process; they are also convenient for systems that intrinsically involve integration. 

\section{Possible Heterodox Individual Origins of the \boldmath$1/f^2$ Photon-Count Spectrum}\label{sec:possorig}

In this heterodox interpretation, the photon counts are modeled as a conditional Poisson process, given \textit{individual} hidden drivers (rates). 
Each source $i$ is thus described by a positive intensity $\mu_i(t)$ governed by its own hidden rate process \(X_i(t)\): 
\begin{equation} \label{muiindiv}
	\mu_i(t)=\mu_{i0}\left[1+\kappa_i X_i(t)\right].
\end{equation}
The driver $X_i(t)$ is taken to have zero mean, with $1+\kappa_i X_i(t)>0$ over the observed interval, and  $\mu_{i0}$ is the mean source intensity. 
Equation~(\ref{muiindiv}) may alternatively be regarded as the leading term in a positive parameterization such as $\mu_i(t)=\mu_{i0}\exp[\kappa_i X_i(t)]$. 
The coupling coefficient $\kappa_i$ relates the magnitude of the source intensity to the hidden rate process. 
All such processes belong to a common class of optical sources for which
\begin{equation} \label{eq:S1f2}
	S_{X_i}(f) \propto \frac{1}{f^2}\,.
\end{equation}
	
As summarized in Sec.~\ref{sec:technoiseexclude}, \(X_i(t)\) cannot plausibly be identified with commonly measured parameters or laboratory variables such as source current, voltage, or temperature; detector noise or temperature; optical-system, environmental, or mechanical noise; nor is it related to device-junction or packaging parameters. Some of these measures may be diagnostic proxies in particular experiments, as discussed in Sec.~\ref{sec:ledsledldincand}, but none can serve as a viable  cause.

Crucially, this approach requires a plausible explanation for the emergence of the inverse-square spectral behavior. Several possibilities, of the many that exist~\cite{KarlinTaylor1981,Metzler2014, Pavliotis2014,GottwaldReich2021}, are considered below. 

	\subsection{Integrated Hidden Forcing}\label{ssec:hiddendynam-intforce}
	
The \(1/f^2\) spectrum suggests that the hidden process $X_i(t)$ may be an integrated stochastic process. 
Consider a generalized integrated process, sometimes termed $1/f^2$ noise or Brownian noise, in which $Y_i(t)$ is defined as a broadband or ``weakly correlated'' forcing function, with
	\begin{equation}
		\frac{dX_i}{dt}=Y_i(t).
	\end{equation}
This leads directly to
	\begin{equation}
		S_{X_i}(f)=\frac{S_{Y_i}(f)}{(2\pi f)^2}, 
	\end{equation}
which, if \(S_{Y_i}(f)\) is approximately flat over the observed band, yields
	\begin{equation}
		S_{X_i}(f) \propto \frac{1}{f^2}\,,
	\end{equation}
as specified in Eq.~(\ref{eq:S1f2}).

If ${Y_i}(t)$ is not strictly uncorrelated, $S_{X_i}(f)$ scales as $1/f^2$ only over the specific frequency band within which $S_{Y_i}(f)$ remains flat.
Although this approach provides a direct mathematical route to the observed spectrum, it is important to emphasize that it does not identify the physical origin of \(Y_i(t)\).	

Standard Brownian motion, a Wiener process, is a particular subset of this formulation. It occurs if and only if ${Y_i}(t)$ is strictly white noise ($\delta$-correlated in time), so that its power spectral density $S_{Y_i}(f)$ is perfectly flat across all frequencies, and Gaussian.
The spectral behavior $S_{X_i}(f) \propto 1/f^2$ is then understood in the finite-band sense relevant to the experimental records. However, the aggregated-photon-count $z$-score traces portrayed in Figs.~\ref{brown}(a)--(e) do not resemble the simulated ordinary Brownian sample paths  plotted in Fig.~\ref{brown}(f). 
		
\subsection{Bounded or Mean-Reverting Diffusion}\label{ssec:hiddendynam-diffbound}
	
A simple mean-reverting diffusion model offers another possibility. Writing
	\begin{equation}
		d{X_i}(t)=-\gamma {X_i}(t)\,dt+\sigma\,dW(t),
	\end{equation}
where $\sigma$ is a constant and \(W(t)\) is standard Brownian motion, leads to the Lorentzian-type spectrum associated with the Ornstein--Uhlenbeck process, 
	\begin{equation}\label{eq:Lorentz}
		S_{X_i}(f)\propto \frac{1}{(2\pi f)^2+\gamma^2}.
	\end{equation}
For frequencies above the relaxation corner,
	\begin{equation}
		f\gg \frac{\gamma}{2\pi}\,,
	\end{equation}
Eq.~(\ref{eq:Lorentz}) can be approximated as 
	\begin{equation}
		S_{X_i}(f) \propto \frac{1}{f^2}\,.
	\end{equation}
This type of model predicts a low-frequency rollover that could lie below the experimentally accessible band.
	
\subsection{Intermittent Diffusion}\label{ssec:hiddendynam-diffintermittent}
	
The final example we consider is intermittent diffusion, 
	\begin{equation}
		d{X_i}(t)=\sigma(t)\,dW(t),
	\end{equation}
in which the parameter \(\sigma(t)\) can vary slowly or intermittently with time. Depending on the character of $\sigma(t)$, this model can generate approximately \(1/f^2\) spectra over finite bands, along with non-Brownian-appearing sample paths that can exhibit:
	\begin{itemize}
		\item Slow drift.
		\item Bursts.
		\item Irregular excursions.
		\item Apparent relaxation-like segments.
	\end{itemize}

The enhanced range of behaviors is phenomenologically attractive for modeling the experimental aggregated-photon-count $z$-score sample paths illustrated in Figs.~\ref{brown}(a)--(e), but it remains crucial to specify the physical origin of the parameters. 
	
\section{Possible Heterodox Common Origins of the \boldmath$1/f^2$ Photon-Count Spectrum}\label{sec:spec}

Again, the photon counts may be modeled as a conditional Poisson process, but now with a \textit{common} hidden external driver rather than individual hidden drivers drawn from a common class, as considered in Sec.~\ref{sec:possorig}.
In this heterodox interpretation, all sources $i$ are governed by the same hidden external varying rate process \(X(t)\):
	\begin{equation}\label{eq:hiddenX}
		\mu_i(t)=\mu_{i0}\left[1+\kappa_i X(t)\right].
	\end{equation}
The driver $X(t)$ is taken to have zero mean, with $1+\kappa_i X(t)>0$, and the $\mu_{i0}$ are the mean source  intensities.
The coupling coefficients $\kappa_i$ relate the magnitudes of the intensities to the rate process and
	\begin{equation}
		S_X(f) \propto \frac{1}{f^2}\,.
	\end{equation}
	
\subsection{Logically Possible Common  Origins}\label{sssec:logical}

Logical possibilities for common hidden external drivers that are ordinarily dismissed as irrelevant or negligible include:
	\begin{itemize}
		\item Geomagnetic and geoelectric field variations.
		\item Atmospheric and ionospheric variations.
		\item Earth-tide and strain-related variations.
		\item Cosmic-ray and secondary-particle flux variations.
		\item Gravitational-wave-background interactions.
		\item Unexplained dark-matter interactions.
	\end{itemize}

The key issue is the identification of a plausible mechanism via which the hidden external driver is sufficiently strongly coupled to the photon-count emission, transmission, or detection probabilities. In all readily identifiable cases, this coupling appears to be too weak by many orders of magnitude, as illustrated by the gravitational-wave-background example provided in Sec.~\ref{sssec:grav}.	

\subsection{Gravitational-Wave Background}\label{sssec:grav}

The gravitational-wave background (GWB) serves as an example of how a physical effect with statistical features that superficially resemble those of the photon count at baseband can fail to account for it by many orders of magnitude in coupling strength.

The theory of general relativity, formulated by Albert Einstein in 1916, revolutionized our understanding of gravity \citep{einstein16a}. One of the most remarkable features of this theory is the prediction that gravitational waves propagating at the speed of light exist~\citep{einstein16b,einstein18}. These ripples in the fabric of spacetime --- characterized by the spacetime strain $h(t)$ --- are generated by accelerating massive objects such as merging black holes or neutron stars~\cite{thorne87}. 
The first direct observation of gravitational waves was achieved in 2015 by the Laser Interferometer Gravitational-Wave Observatory (LIGO), whose original  sites were at Livingston, Louisiana and Hanford, Washington~\citep{abbott16,abac2026}.

The stochastic GWB arises from the unresolved superposition of gravitational waves from large numbers of individual black hole binaries across the universe. 
In the frequency range \(10^{-6}\) to \(10^{-3}\)~Hz, the GWB is thought to arise from merging supermassive black hole (SMBH) binaries at different stages of their inspiral (which typically last years to decades).
The GWB can be characterized as a ``hum,'' in contrast to the ``shouts'' of individual black-hole mergers. 

The standard GWB superposition model~\cite{phinney01,farmer03,rosado11,	sesana13a,sesana13b} leads to a simple and robust spectral form for the strain that has its origin in the deterministic inspiral spectra of the constituent individual binaries. 
In the context of the current discussion, the most useful spectral form that characterizes the GWB is the one-sided strain power spectral density
\begin{equation}
	S_h(f) \propto f^{-7/3},
	\label{eq:sh_fminus_seventhirds}
\end{equation}
which is close to the observed $1/f^2$ one-sided (baseband) photon-count spectral density.
Other commonly used GWB spectral forms that are equivalent to Eq.~(\ref{eq:sh_fminus_seventhirds}) 
are the strain amplitude per logarithmic frequency interval
$h_c(f) \propto f^{-2/3}$, and the dimensionless gravitational-wave energy-density spectrum
$\Omega_{\scriptscriptstyle \rm GW}(f) \propto f^{2/3}$.

Several statistical features of the low-frequency photon count resemble those of the GWB: 1)~the spectral slope: $S_P(f) \propto f^{-2} \Leftrightarrow S_h(f) \propto f^{-7/3}$;\, 2)~the frequency range of the fluctuations: $1 \times 10^{-6} \leqslant f \leqslant 5 \times 10^{-4}~\text{Hz} \Leftrightarrow 10^{-6} \leqslant f \leqslant 10^{-3}~\text{Hz}$; and 3)~the highly irregular $z$-score sample paths. 

If we assume that the gravitational-wave spacetime strain \(h(t)\) represents the common external rate process \(X(t)\) that governs the intensity $\mu(t)$,  Eq.~(\ref{eq:hiddenX}) can be cast in the form
\begin{equation}\label{eq:hiddenXepsilon}
	\frac{\delta \mu(t)}{\mu_0} \approx \kappa h(t).
\end{equation}
Adopting a commonly used order-of-magnitude estimate for the gravitational-wave strain amplitude,
\begin{equation}
   10^{-18}	\lesssim  h \lesssim 10^{-15}, 
\end{equation}
and assuming that fractional intensity fluctuations $\delta \mu/\mu_0 \approx 10^{-3}$ suffice for observation, yields a required coupling coefficient in the range
	\begin{equation}
		10^{12}	\lesssim  \kappa \lesssim 10^{15}.
	\end{equation} 
	
In short, despite the similarities in the statistical features of the photon count at baseband and the GWB, standard gravitational coupling is some 12--15 orders of magnitude too weak to explain the $1/f^2$ photon-count spectrum at low frequencies.
If the common external process \(X(t)\) were time dilation rather than the gravitational wave strain itself, the required coupling coefficient would have to be even larger.

\section{Suggested Related Experiments}\label{suggest}

Finally, we suggest two additional classes of experiments that could further elucidate the origin of the $1/f^2$ photon-count fluctuations at baseband. 

\subsection{Independent Optical Sources}\label{indepsources}

It may be possible to establish whether individual or common hidden origins are at play in the photon-count spectra, as described in Secs.~\ref{sec:possorig} and \ref{sec:spec}, respectively, by simultaneously carrying out experiments that use two independent and co-located optical sources with separate optics and identical photodetectors. If the same underlying rate function \(X(t)\) drives the slow photon-count fluctuations across different sources, simultaneously measured independent sources should exhibit correlated long-timescale fluctuations.

\subsection{Ionizing Radiation and Solid-State  Detectors}\label{gammaphotons}

Experiments analogous to those reported in Sec.~\ref{Exp}, which make use of optical sources and photodetectors, can be carried out using sources of ionizing radiation and direct-conversion solid-state radiation detectors~\cite{knoll2010,grupen2016,tsoulfanidis2021}. These experiments would not make use of light and would contain no optical components, so they would be useful in assessing the precise role played by light and optics in generating the $1/f^2$ baseband spectrum. 

A direct carryover from the optical- to the ionizing-radiation domain would entail measuring the low-frequency spectra of the output counts at a pair of solid-state radiation detectors (e.g., bulk CdZnTe) irradiated by two independent and co-located $\gamma$-particle sources (e.g., $^{137}$Cs), in place of the output counts at a pair of SPADs or PMTs illuminated by two independent and co-located optical-photon sources.

The same approach could be used to examine 
the low-frequency particle-count spectra associated with $\beta^-$- or $\alpha$-particles. Those experiments would make use of a pair of Si $p$--$i$--$n$ radiation detectors irradiated by two independent and co-located $\beta^-$ or $\alpha$ emitters (e.g., $^{90}$Sr$/^{90}$Y or  $^{241}$Am, respectively). 
Whereas the $\gamma$-particle (photon) is a massless elementary spin-1 boson, the $\beta^-$-particle (electron) is a massive elementary spin-$\tfrac{1}{2}$ fermion, and the $\alpha$-particle (${}^4$He nucleus) is a massive composite spin-0 boson.
It would be of interest to establish whether there is a linkage between the quantum statistics of the particles being counted and the presence of low-frequency particle-count fluctuations.

It would also be instructive to compare the results of the betaluminescence experiments described in Sec.~\ref{sec:betalum}, which relied on the conversion of $\beta^-$-particles to optical luminescence photons and made use of optical photon counting, with the results of direct-conversion experiments in which the $\beta^-$-particles are directly converted and counted without optical mediation. 

\section{Conclusion}\label{concl}
	
Our investigation has narrowed the conventional explanatory space that might account for the presence of universal $1/f^2$ photon-count fluctuations at baseband. 	
The available data point to photon counts that obey a doubly stochastic Poisson counting process with a slowly varying random intensity.

The presence of these fluctuations in betaluminescence  is especially elucidating because the photon generation process is not driven by an applied current or voltage, but rather by nuclear decay that is  temperature independent and has an essentially constant rate.
	
The physical origin of the intensity fluctuations remains unknown, but likely falls within one of two broad possibilities:
	\begin{itemize}
		\item \textit{Individual hidden drivers} $X_i(t)$, whereby different sources are governed by their own independent slow random rate fluctuations, but all belong to the same broad dynamical class that naturally produces $1/f^2$ spectra.
		\item \textit{A common external driver} $X(t)$, whereby a real but unidentified process slowly modulates the photon-count rates of a collection of different sources.
	\end{itemize}
Additional experiments are suggested to further clarify the origin of these $1/f^2$ fluctuations.
	
\vspace{-1mm}
\begin{acknowledgments}
	We are grateful to Boston University and the Boston University Photonics Center for generous financial and logistical support. 
\end{acknowledgments}

\bibliography{constraints}

\end{document}